\pdfoutput=1
\documentclass{JINST}

\usepackage{graphicx}
\usepackage{bm}
\newcommand{\ua}{\uparrow}
\newcommand{\da}{\downarrow}

\newcommand{\avrg}[1] { \left< #1 \right> }
\title{Improved Method to extract Nucleon Helicity Distributions using 
Event Weighting}

\author{J\"org Pretz$^{a,b,c}$\\
\llap{$^a$}III. Physikalisches Institut B, RWTH Aachen University,
Otto-Blumenthal-Stra{\ss}e, 52056 Aachen, Germany\\
\llap{$^b$}Institut f\"ur Kernphysik, Forschungszentrum J\"ulich GmbH,
Wilhelm-Johnen-Stra{\ss}e, 52428 J\"ulich, Germany\\

E-mail: \email{pretz@physik.rwth-aachen.de}
}

 \abstract{
An improved analysis method to extract quark helicity distributions in leading
order (LO) QCD
from semi-inclusive double spin asymmetries in deep inelastic scattering is presented.
The method relies on the fact that fragmentation functions, describing the fragmentation
of a quark into a hadron, have a strong dependence on the energy fraction $z$ of the observed hadron.
Hadrons with large $z$ contain more information about the struck quark. This can be used in a weighting
procedure to improve the figure of merit (= inverse of variance).
In numerical examples it is shown that one could gain 15-39\% depending on the quark flavor and cut on $z$.
 
Mathematically the problem can be described as finding an optimal solution
 in terms of the figure of merit for parameters $\boldsymbol \Theta$ determined from a system of linear equations 
 ${\bf B}(x) {\boldsymbol \Theta} ={\bf Y}(x)$,
where the measured input vector ${\bf Y}(x)$ is given as  event distributions depending on a random variable $x$, 
the coefficients of the matrix ${\bf B}(x)$ depend as well on $x$, whereas the  parameter vector $\boldsymbol\Theta$ to be 
 determined does not. 
 }

\keywords{data analysis, likelihood, asymmetry, event weighting,
helicity distributions, fragmentation functions}

\begin{document}

 \section{Introduction}
Consider an event distribution depending on a random variable $x$ of the following form:
 \[
 n^\pm(x) = \alpha(x) (1 \pm \beta(x) P) 
  \]
  with the goal to extract $P$ with the highest precision.
  In many experimental situations only $\beta(x)$  is known, whereas $\alpha(x)$ contains
  not very precisely known flux and acceptance factors.
One concrete example is the determination of a polarization $P$ from event rates in two different
polarization configurations, $n^\pm(x)$, measured as a function of the polar angle $x$ with known analyzing power $\beta(x)$
\cite{Fernow:1981fw}.  
  
One straightforward way to extract $P$ is to integrate $n^\pm(x)$ over $x$
to obtain the total number of events. The expectation value is given by
\begin{eqnarray}
\avrg{n^\pm}  &=& \int \alpha(x) dx  \pm P \int \alpha(x) \beta(x) dx\\
              &=& (1 \pm P \avrg{\beta}) \,\int \alpha(x) dx
\end{eqnarray}
with $\avrg{\beta} = \int \alpha \beta dx/\int \alpha dx$.
The asymmetry 
\[
  \frac{\avrg{n^+}-\avrg{n^-}}{\avrg{n^+}+\avrg{n^-}} = \avrg{\beta} P 
\]
gives direct access to $P$.
This leads to the following estimator $\hat P$ for $P$:
\begin{equation}
 \hat P = \frac{N^+ - N^-}{\sum_+ \beta(x_i) + \sum_- \beta(x_i)} \, ,
\end{equation}
where $N^\pm$ are the number of observed events.
The two sums run over all observed events in the given configuration.

For small asymmetries $(N^+ \approx N^-)$ the figure of merit (FOM) is given
by (see ref.~\cite{Pretz:2011qb})
\begin{eqnarray}
 \mbox{FOM} &=& (\avrg{n^+} + \avrg{n^-}) \avrg{\beta}^2 \\
           &=& (N^+ + N^-) \overline{\beta}^2 \, .
\end{eqnarray}
The second equation indicates, that to evaluate the FOM, the expectation values 
are replaced by the actual number of observed events $N^\pm$
and 
\[
\avrg{\beta} \approx \overline{\beta} =  \frac{\sum_+ \beta(x_i) + \sum_- \beta(x_i)}{N^+ + N^-} \, .
\]
In the following we assume that the data sample is large enough so that the distinction
between expectation values  and observed number of events or averages over event samples
does not matter.

In ref.~\cite{Pretz:2011qb} is was shown that 
assigning to every event a weight factor equal to its analyzing power $\beta(x_i)$ 
one reaches a larger FOM.
The estimator in this case is
\begin{equation}\label{alh}
 \hat P = \frac{\sum_+ \beta(x_i) - \sum_- \beta (x_i)}{\sum_+ \beta^2(x_i) +
   \sum_- \beta^2(x_i)} \, 
\end{equation}
and the FOM reads
\begin{equation}
 \mbox{FOM} = (N^+ + N^-) \avrg{\beta^2} \, ,
\end{equation}
thus gaining a factor $\avrg{\beta^2}/\avrg{\beta}^2$. 
It is evident that the gain is the larger the stronger is the $x$-dependence of $\beta(x)$.

In this work the problem will be extended to the case where several parameters $P_1, \dots P_n$ have to be determined
from $2m$ event rates $n_i^{\pm}, i=1, \dots m$.
This gives the following $2m$ equations: 
 \begin{eqnarray}
  n^\pm_1(x) &=& \alpha_1(x) \left(1\pm \beta_{11}(x) P_1 \pm \beta_{12}(x)
  P_2 \pm \dots \beta_{1n}(x) P_n \right) \, ,\\
  n^\pm_2(x) &=& \alpha_2(x) \left(1\pm\beta_{21}(x) P_1 \pm \beta_{22}(x)
  P_2 \pm \dots \beta_{2n}(x) P_n  \right) \, ,\\
         & &  \vdots \nonumber \\
  n^\pm_m(x) &=& \alpha_m(x) \left(1\pm\beta_{m1}(x) P_1 \pm \beta_{m2}(x)
  P_2 \pm \dots \beta_{mn}(x) P_n \right) \, .
 \end{eqnarray}
For $m=1$ and $n=2$ this has been discussed in ref.~\cite{Pretz:2008mi}
and applied in refs.~\cite{Alekseev:2008cz} and \cite{Adolph:2015cvj}. 

The paper is organized as follows.
In section~\ref{poldis} the relation between event rates measured in deep inelastic scattering 
and polarized quark distributions will be discussed.
Section~\ref{methods} compares different methods to extract polarized quark distributions from data.

\section{Double spin asymmetries in deep inelastic scattering}\label{poldis}
Up to now the discussion was rather academic. As a concrete example consider double spin asymmetries $A^h$ measured in  
semi-inclusive deep inelastic scattering~\cite{Alekseev:2010ub}:
\[
\vec{\ell}(k) + \vec{N}(p) \rightarrow \ell'(k') + h(p_h) + X(p_X) \, .
\]
They give access
to quark helicity distributions $\Delta q$ of the nucleon. A lepton $\ell$ is scattered off a nucleon $N$.
In the final state the lepton $\ell'$ and one or more hadrons $h$ are observed. The variables in parentheses
denote the four-vectors of the particles.
The helicity distributions $\Delta q =  q^\ua- q^\da$ are defined as the difference of number of quarks of flavor $q$ with spin aligned to the nucleon spin $(q^\ua)$
minus the number of quarks with spin anti-aligned to the nucleon spin $(q^\da)$.

The number of hadrons of species $h$ detected with target and beam polarization parallel ($\ua \ua$) or 
antiparallel ($\ua \da$) is  related to the asymmetries $A^h$ by
\begin{equation}
 n_h^{\ua \da (\ua \ua)}(x,z,Q^2) = \alpha(x,z,Q^2) (1 \pm \beta(x,z,Q^2) A^h(x,z,Q^2))  
\end{equation}
with
\begin{eqnarray}
 \alpha &=& \Phi a n \sigma \, ,\\
 \beta &=& f P_T P_B D \, .\label{eq:beta}
\end{eqnarray}
$ \Phi, a, n, \sigma$ are flux, acceptance, target density and unpolarized cross section~\cite{Alekseev:2010ub}, respectively.
Since we consider only asymmetries, these factors drop out of the following equations.
The factors $f, P_T, P_B, D$ denote the target dilution factor, target polarization, beam polarization and depolarization factor~\cite{Alekseev:2010ub},
respectively. For the discussion here these factors are set to unity to simplify the notation.

In leading-order QCD the double spin virtual photon 
 asymmetries $A^h$ are related in the following way to the quark helicity distributions $\Delta q$:
 \begin{equation}\label{eq:ah}
  A^h(x,z,Q^2) = \frac{\sum_{q} e_q^2 \Delta q(x,Q^2) D_q^h(z,Q^2)}{\sum_{q} e_q^2 q(x,Q^2) D_q^h(z,Q^2)} \, .
 \end{equation}
 The sums run over the quark flavors $q=u,d,s,\bar u, \bar d, \bar s$.
 The variables $x,z$ and $Q^2$ have their usual meanings in deep inelastic scattering (see Tab.~\ref{tab:vari}).
 %
 \begin{table}
\begin{center}
  \begin{tabular}{|l|l|}
  \hline
  Variable & Meaning \\
 \hline
 $Q^2 = - q^2 = (k-k')^2$ & 4-momentum transfer\\
 $x = \frac{Q^2}{2 p q}$ & Bjorken variable\\
 $z = \frac{p_h p}{q p}$ & energy fraction of virtual photon taken \\
                       & by observed hadron in target rest frame\\
 \hline
  \end{tabular}
\end{center}
\caption{Kinematic variables in deep inelastic scattering. \label{tab:vari}}
 \end{table}
 $q(x,Q^2)$ are the unpolarized parton distributions and $e_q$ are the  charges of the quarks.
 $D_q^h(z)$ are the fragmentation functions related to the probability that a struck quark $q$ 
 fragments into a hadron $h$ with energy fraction $z$.
 In the analysis of ref.~\cite{Alekseev:2010ub} the event rates are integrated over $z$ and $Q^2$
 and the helicity distributions are obtained for every bin in $x$ independently
 using the semi-inclusive asymmetries of eq.~(\ref{eq:ah})
 for different hadron species and inclusive asymmetries on different targets
 solving a system of linear equations.
 
 Aim of this paper is to show that by integrating over $z$ one loses information.
 It is shown that using a different analysis scheme, outlined below, one could extract the helicity distributions with
 higher accuracy.
 The reason lies in the form of the fragmentation functions.
 They have a strong $z$ dependence.

\begin{figure}
  \includegraphics[width=\textwidth]{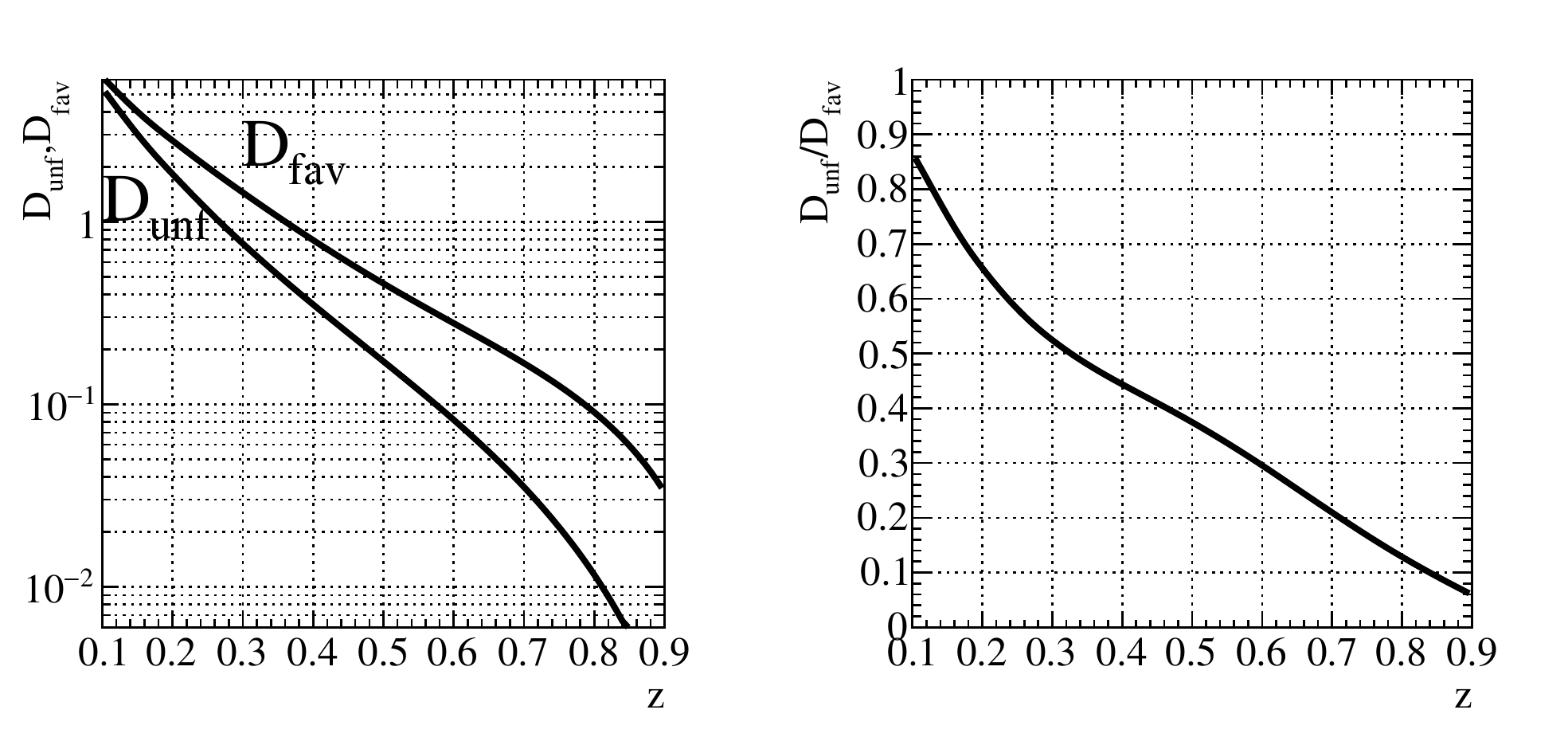}
  \caption{The fragmentation functions $D_{fav} = D_u^{\pi^+} = D_{d}^{\pi^-}$  and  $D_{unf} = D_{u}^{\pi^-} = D_{d}^{\pi^+}$
  from ref.~\cite{deFlorian:2007aj} in LO QCD at $Q^2=5$GeV$^2$  (left) and their ratio (right). \label{fig:ff}}
 \end{figure}

 To understand the principle, we consider the simplest case of a proton target with just $u$ and $d$ quarks and 
 asymmetries for positively and negatively charged pions. In this case only two different fragmentation functions are
 involved denoted as favored, $D_{fav} = D_u^{\pi^+} = D_{d}^{\pi^-}$, and unfavored, $D_{unf} = D_{u}^{\pi^-} = D_{d}^{\pi^+}$,
 depending on whether the struck quark is a valence quark of the observed hadron or not.
 Further, as mentioned above, we set all experimental factors to unity and consider a fixed $x$ and assume $u=2$ and $d=1$
 for a proton target.
 We then end up with the following equations (using now the notation $+$ for $\pi^+$ and $-$ for $\pi^-$).
 \begin{eqnarray}
   n_{\ua \da}^{+}(z) &\propto & \alpha^{+} (1 +  \beta^{+}_u(z) \Delta u +
   \beta^{+}_d (z) \Delta d) \, ,\label{nplpl}\\  
   n_{\ua \ua}^{+}(z) &\propto & \alpha^{+} (1 -  \beta^{+}_u(z) \Delta u -
   \beta^{+}_d (z) \Delta d) \, ,\\
   n_{\ua \da}^{-}(z) &\propto & \alpha^{-} (1 +  \beta^{-}_u(z) \Delta u +
   \beta^{-}_d (z) \Delta d) \, ,\\  
   n_{\ua \ua}^{-}(z) &\propto & \alpha^{-} (1 -  \beta^{-}_u(z) \Delta u -
   \beta^{-}_d (z) \Delta d) \label{nmimi}
 \end{eqnarray}
 with 
 \begin{eqnarray}
  \alpha^{+} = \frac{4}{9} u D_{fav} + \frac{1}{9} d D_{unf} \, ,&&
  \alpha^{-} = \frac{4}{9} u D_{unf} + \frac{1}{9} d D_{fav} \, ,\\
                 & & \nonumber \\
  \beta_u^{+}  = \frac{4 D_{fav}}{4 u D_{fav} +  d D_{unf}} \, ,&& 
  \beta_d^{+}  = \frac{D_{unf}}{4 u D_{fav} +  d D_{unf}} \, ,\\ 
 \beta_u^{-}  = \frac{4 D_{unf}}{4 u D_{unf} +  d D_{fav}} \, ,&&
  \beta_d^{-}  = \frac{D_{fav}}{4 u D_{unf} +  d D_{fav}} \, .
 \end{eqnarray}
 The fragmentation functions $D_{fav}$ and $D_{unf}$ and their ratio are shown in Fig.~\ref{fig:ff} 
 using a LO parameterisation at $Q^2=5$GeV$^2$ of ref.~\cite{deFlorian:2007aj,ffgen}. This figure indicates that an event at large $z$ should get 
 in general a larger
 weight in the analysis, since it carries more information on the struck quark. At $z\approx0.7$ it is five times more likely
 that a $\pi^+$ originated from a $u$ quark than from a $d$ quark. At $z\approx 0.1$ these probabilities are almost equal.
 
 \section{Different methods to extract the quark helicity distributions}\label{methods}
 In this section several methods to determine the quark helicity distributions
 are discussed, starting with the counting rate method.
 Then the unbinned maximum likelihood (MLH) method known to give
 the largest FOM is discussed, followed by the newly proposed weighting method, shown to give the same FOM
 as the maximum likelihood method.
 Finally, a method using bins in $z$ is discussed.
 
 \subsection{Counting rate asymmetry}
 
 Integrating eqs.~(\ref{nplpl}) -(\ref{nmimi}) over $z$ and then
 forming asymmetries for $\pi^+$ and $\pi^-$ leads to the following two equations
 \begin{eqnarray}
  A^+ &=& \avrg{\beta_u^+} \Delta u + \avrg{\beta_d^+} \Delta d \, , \\
  A^- &=& \avrg{\beta_u^-} \Delta u + \avrg{\beta_d^-} \Delta d 
 \end{eqnarray}
with 
\[
 \avrg{\beta_q^c} = \frac{\int \alpha^c \beta_q^c dz }{\int \alpha^c dz}
 \approx \frac{\sum_{\ua \da} \beta_q^c(z_i)+ \sum_{\ua \ua} \beta_q^c(z_i)}{N^{\ua \da}
 + N^{\ua \ua}} \, ,\quad  q=u,d, \quad c=+,- \, .
 \]
The covariance matrix for $\Delta u$ and $\Delta d$ reads
 \[
\mbox{cov}(\Delta u,\Delta d)^{-1} =   {\bf B}^T \mbox{cov}(A^+,A^-)^{-1} {\bf
 B}
 \]
 with
 \[
  {\bf B} = \left(
  \begin{array}{cc}
   \avrg{\beta_u^{+}} & \avrg{\beta_d^{+}} \\ 
   \avrg{\beta_u^{-}} & \avrg{\beta_d^{-}} \\ 
  \end{array} 
  \right)
 \]
and 
\[
 \mbox{cov}(A^+,A^-) =
 \left(
 \begin{array}{cc}
  1/N^+ & 0 \\
  0  &  1/N^- \\
 \end{array}
 \right) \, ,
\]
assuming Poisson distributed events and $A^{\pm}\ll 1$.
Note that the assumption $A^{\pm}\ll 1$ (or equivalently $\beta \Delta q \ll 1$) is not really a restriction because
in reality this is always fulfilled, since $\beta \approx 0.1$ if one includes
the experimental factors $f P_T P_B D$ of eq.~(\ref{eq:beta}).

This results in the following FOM for $\Delta u$ (a similar expression can be obtained for $\Delta d$):
 \begin{equation}\label{fom_u_cr}
  \mbox{FOM}_{\Delta u} = \left(N^+ \avrg{\beta_u^{+}}^2 + N^- \avrg{\beta_u^{-}}^2 \right) (1-\rho^2) \label{eq:fom_w1}
 \end{equation}
with
 \[
  \rho = - \frac{ \left(N^+ \avrg{\beta_u^{+}} \avrg{\beta_d^{+}} +N^- \avrg{\beta_u^{-}}\avrg{\beta_d^{-}} \right) }
  {\sqrt{(N^+ \avrg{\beta_u^{+}}^2 + N^-  \avrg{\beta_u^{-}}^2)
 (N^+ \avrg{\beta_d^{+}}^2 + N^-  \avrg{\beta_d^{-}}^2) }  } \, ,
 \]
where $\rho$ is the correlation between $\Delta u$ and $\Delta d$.

 \subsection{Maximum Likelihood Method}
 Statistically the most efficient method is the unbinned maximum likelihood method .
 Applied to the example above, one finds for the log-likelihood function
 \begin{eqnarray}
 \ell = \log \mathcal{L} & = & \sum_{\ua \da} \log (\alpha^{+}(z_i) (1 +  \beta^{+}_u(z_i) \Delta u + \beta^{+}_d (z_i) \Delta d)) 
               - \avrg{n_{\ua \da}^{+}} \nonumber \\
 & + & \sum_{\ua \ua} \log (\alpha^{+}(z_i) (1 -  \beta^{+}_u(z_i) \Delta u - \beta^{+}_d (z_i) \Delta d)) - \avrg{n_{\ua \ua}^{+}} \nonumber \\
 & + & \sum_{\ua \da} \log (\alpha^{-}(z_i) (1 +  \beta^{-}_u(z_i) \Delta u + \beta^{-}_d (z_i) \Delta d)) - \avrg{n_{\ua \da}^{-}} \nonumber \\  
  & + &  \sum_{\ua \ua} \log( \alpha^{-}(z_i) (1 -  \beta^{-}_u(z_i) \Delta u
 - \beta^{-}_d (z_i) \Delta d))- \avrg{n_{\ua \ua}^{-}} \, .
 \end{eqnarray}

 The sums run over all observed hadrons in the considered polarization configuration. 
 Terms like  $\avrg{n^{+}_{\ua\ua}}$ are included 
 because the extended MLH~\cite{Barlow:213033} has to be applied.

 The problem can easily be solved numerically by solving the two equations $\partial \ell/\partial \Delta u = 0$ and $\partial \ell/\partial \Delta d=0$
 for $\Delta u$ and $\Delta d$. For $\beta \Delta q \ll 1, (\Delta q = \Delta u, \Delta d)$ even an analytic solution exists.
 Here we are more interested in the covariance matrix. In general it is given by
 \begin{eqnarray}
  \mbox{cov}^{-1}(\Delta u, \Delta d)  &=&  
  - \left(
  \begin{array}{cc}
   \frac{\partial^2 \ell}{\partial \Delta u^2}
  &  \frac{\partial^2 \ell}{\partial \Delta u \partial \Delta d} \\
   \frac{\partial^2 \ell}{\partial \Delta u \partial \Delta d} & \frac{\partial^2 \ell}{\partial \Delta d^2} \\
  \end{array}
\right)  \, .
\end{eqnarray}
For $\beta \Delta q \ll 1$ one finds
\begin{eqnarray}
  \mbox{cov}^{-1}(\Delta u, \Delta d)  &=&  
\left(
  \begin{array}{cc}
   \sum (\beta^{+}_u(z_i))^2 + (\beta^{-}_u(z_i))^2 & \sum \beta^{+}_u(z_i) \beta^{+}_d(z_i) + \beta^{-}_u(z_i) \beta^{-}_d(z_i) \\
   \sum \beta^{+}_u(z_i) \beta^{+}_d(z_i) + \beta^{-}_u(z_i) \beta^{-}_d(z_i) & \sum (\beta^{+}_d(z_i))^2 + (\beta^{-}_d(z_i))^2 \\
  \end{array}
\right) \nonumber \\
&=&
\left(
  \begin{array}{cc}
   N^+ \avrg{(\beta^{+}_u)^2} + N^- \avrg{(\beta^{-}_u)^2} \, & N^+\avrg{\beta^{+}_u \beta^{+}_d} + N^- \avrg{\beta^{-}_u \beta^{-}_d} \\
   N^+\avrg{\beta^{+}_u \beta^{+}_d} + N^- \avrg{\beta^{-}_u \beta^{-}_d} \, &
   N^+ \avrg{(\beta^{+}_d)^2} + N^- \avrg{(\beta^{-}_d)^2} \\
  \end{array}
\right) \, .\label{cov_dq_mlh}
 \end{eqnarray}
The sums run over both polarisation configurations ($\ua\da$ and $\ua\ua$).

The figure of merit FOM of $\Delta u$ is found to be
\begin{equation}\label{fom_u_lh}
 \mbox{FOM}_{\Delta u} = \left( N^+ \avrg{(\beta^{+}_u)^2} + N^- \avrg{(\beta^{-}_u)^2} \right) \,  (1-\rho^2) \label{eq:fom_mlh}
\end{equation}
with 
\[
 \rho = - \frac{N^+\avrg{\beta^{+}_u \beta^{+}_d} + N^- \avrg{\beta^{-}_u \beta^{-}_d}}
 {\sqrt{ \left(N^+ \avrg{(\beta^{+}_u)^2} + N^- \avrg{(\beta^{-}_{u})^2} \right) \, \left(  N^+ \avrg{(\beta^{+}_d)^2} + N^- \avrg{(\beta^{-}_d)^2
 \right)}        }}
\]
being the correlation between $\Delta u$ and $\Delta d$.
Equation~(\ref{fom_u_lh}) is similar to eq.~(\ref{fom_u_cr}).
For the likelihood method factors $<\beta ^2>$ occur instead of $<\beta>^2$.

\subsection{Weighting Method}
\begin{figure}
 \includegraphics[width=\textwidth]{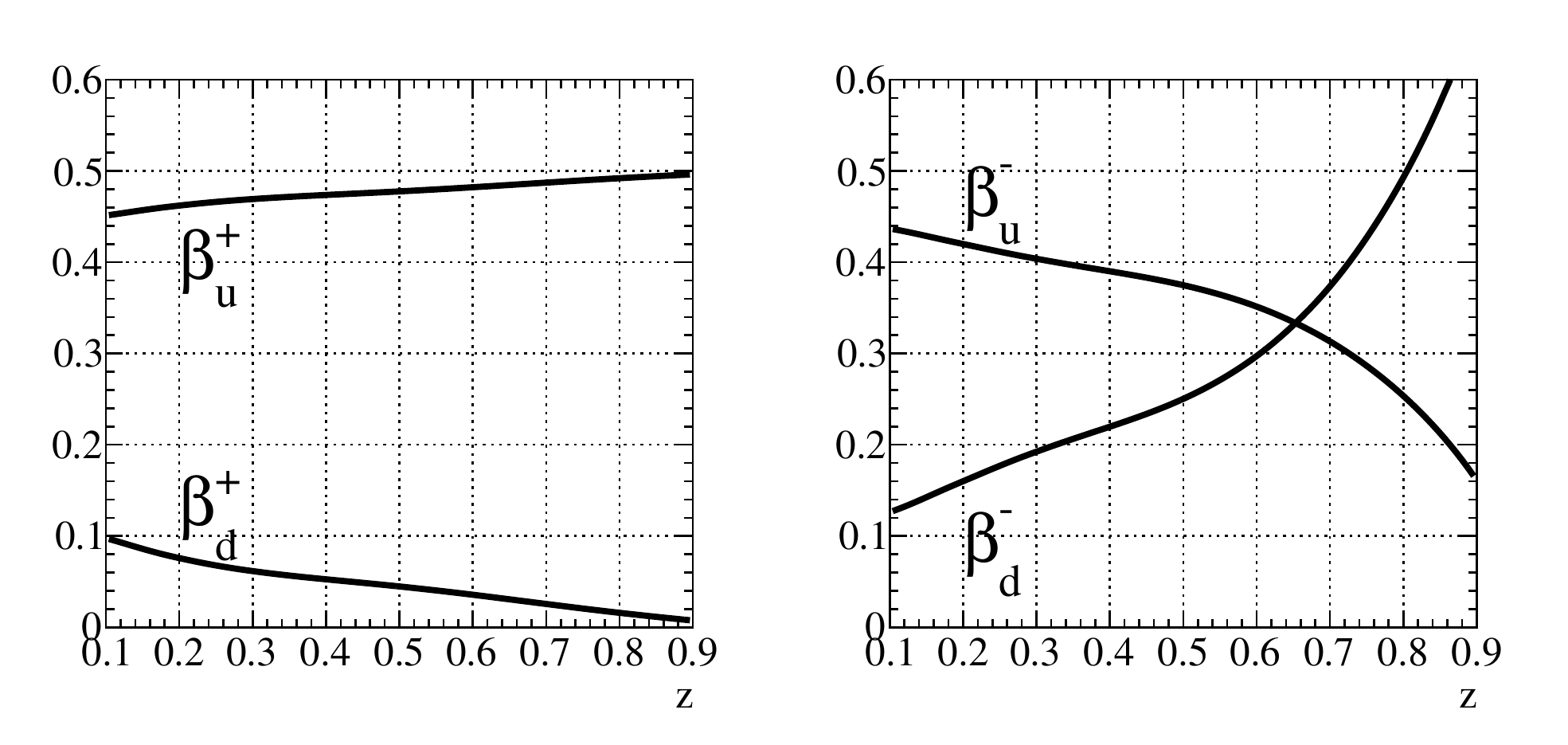}
 \caption{ $\beta_u^{+}, \beta_d^{+}$ (left) and $\beta_u^{-}, \beta_d^{-}$ (right) vs. $z$. \label{fig:beta}}
\end{figure}

It is less convenient to work with the unbinned maximum likelihood method because it involves 
sums over all events in the minimization process. A way out is to consider asymmetries of weighted events.
Guided by the example with one unknown, one considers event weighted asymmetries 
of the type:
\[
 \hat a_{\beta_u^{+}} := \frac{ \sum_{\ua \da} \beta_u^{+}(z_i) - \sum_{\ua \ua} \beta_u^{+}(z_i)}
 {\sum_{\ua \da} (\beta_u^{+}(z_i))^2 +\sum_{\ua \ua} (\beta_u^{+}(z_i))^2}\, ,
\]
i.e. the weight factor is given by the $\beta$-factor in front of the quantities $\Delta u, \Delta d$ in eqs.~(\ref{nplpl})-(\ref{nmimi}).
The $z$ dependence of the $\beta$s, which determine the gain in weight, are mainly determined by the ratio of fragmentation functions.
The four $\beta$ factors are shown in Fig.~\ref{fig:beta}.

Finally, this leads to the following system of linear equations
\begin{eqnarray}
\avrg{ \hat a_{\beta_u^{+}}} &=& \Delta u+  \frac{\avrg{ \beta_u^{+} \beta_d^{+}}}{\avrg{ (\beta_u^{+})^2}} \, \Delta d \, ,\label{awplu}\\
 \avrg{\hat a_{\beta_d^{+}}} &=&   \frac{\avrg{ \beta_u^{+} \beta_d^+}}{\avrg{ (\beta_d^{+})^2}}\, \Delta u +  \Delta d \, ,\\
\avrg{\hat a_{\beta_u^{-}}} &=&  \Delta u+  \frac{\avrg{ \beta_u^{-} \beta_d^{-}}}{\avrg{ (\beta_u^{-})^2}} \, \Delta d \, , \\
 \avrg{\hat a_{\beta_d^{-}}} &=&  \frac{\avrg{ \beta_u^{-} \beta_d^-}}{\avrg{ (\beta_d^{-})^2}} \, \Delta u+  \Delta d \, .\label{awmid}
 \end{eqnarray}
 This can be written in matrix from as
 \[
  {\bf a = B }\Delta {\bf q}
 \]
with ${\bf a} =
(\hat a_{\beta_u^{+}},\hat a_{\beta_d^{+}},\hat a_{\beta_u^{-}},\hat a_{\beta_d^{-}})^T$, ${\Delta \bf q} = (\Delta u, \Delta d)^T$
and $\bf B$ a $4 \times 2$ matrix with coefficients given by the factors in front of $\Delta u$ and $\Delta d$ in eqs.~(\ref{awplu}) - (\ref{awmid}).
In general, for each quark flavor to be determined a weighted asymmetry per
hadron species has to be evaluated.
For the case considered here
this gives in total 4 asymmetries (2 quark flavors $\times$ 2 hadron species).

The covariance matrix for the weighted asymmetries is given by (see appendix~\ref{app1}) 
\begin{equation}\label{cov_aw}
 \mbox{cov}({\bf a}) =
 \left(
 \begin{array}{cccc}
  \frac{1}{N^{+} \avrg{(\beta_u^{+})^2}}  & \frac{\avrg{\beta_u^{+}
      \beta_d^+}}{N^{+} \avrg{(\beta_u^{+})^2} \avrg{(\beta_d^{+})^2}}  &  0& 0 \\ 
 \frac{\avrg{\beta_u^{+} \beta_d^+}}{N^{+} \avrg{(\beta_u^{+})^2 }\avrg{(\beta_d^{+})^2}} & \frac{1}{N^{+} \avrg{(\beta_d^{+})^2} } &   0& 0 \\ 
 0 & 0 & \frac{1}{N^{-} \avrg{(\beta_u^{-})^2}} &   \frac{\avrg{\beta_u^{-} \beta_d^-}}{N^{-} \avrg{(\beta_u^{-})^2 }\avrg{(\beta_d^{-})^2}}                    \\
 0 & 0 &  \frac{\avrg{\beta_u^{-} \beta_d^-}}{N^{-} \avrg{(\beta_u^{-})^2 }\avrg{(\beta_d^{-})^2}}  & \frac{1}{N^{-} \avrg{(\beta_d^{-})^2}} \\
 \end{array}
\right) \, . 
\end{equation}
Note, that no correlation between different hadron species (here $+$ and $-$) is considered here.
It can easily be implemented in the weighting method (see appendix~\ref{app1}), but not so easily in the
MLH method.

In this way one obtains a system of 4 linear equations with 2 unknowns which can be solved be least squares minimization.
The covariance matrix of $\Delta u$ and $\Delta d$ is given by
\begin{equation}\label{cov_w}
 \mbox{cov}(\Delta u, \Delta d)^{-1} = {\bf B}^T  \mbox{cov}({\bf a})^{-1} {\bf B} \, .
\end{equation} 
Performing the matrix multiplication in eq.~(\ref{cov_w}) shows that $\mbox{cov}(\Delta u, \Delta d)$ 
equals the covariance matrix found in the MLH method.

Thus the weighting method allows one to extract the polarized quark distributions with the same uncertainty
as the MLH method but with a higher FOM as the counting 
rate method used in ref.~\cite{Alekseev:2010ub}.
~\footnote{Note however that 
	ref.~\cite{Alekseev:2010ub} used a weighting procedure, but only in the factor
	$f P_T P_B D$ in eq.~(\ref{eq:beta})}

\subsection{Binning data in $z$}
The discussion showed that using event weighting leads to the maximal FOM.
However, often experimental asymmetries are used by theorists to perform global analysis
of the several experiments going beyond the LO QCD discussed here~\cite{Leader:2010rb,Arbabifar:2013tma}.

The disadvantage for experimental groups to publish weighted asymmetries is that they depend on a choice
of fragmentation functions and unpolarized quark distributions. Counting rate asymmetries (up to a small dependence due to the dilution factor
and radiative corrections) do not.
One way out is to publish in addition counting rate asymmetries in bins of $z$.
In this subsection it is shown that in the limit of infinitely small bins in
$z$ one reaches the FOM of the MLH method as well.
Of course a fine binning is limited in practice by the fact
that mainly at large $z$ data are sparse.

Using $n$ bins in $z$, one obtains the following system of equations:
\begin{eqnarray}
 \left(
 \begin{array}{c}
  A_1^+ \\
  \vdots \\
  A_n^+\\
   A_1^- \\
   \vdots \\
   A_n^- \\
 \end{array}
 \right) 
 =
 \left(
 \begin{array}{cc}
 \avrg{ \beta_{u,1}^+} & \avrg{\beta_{d,1}^+} \\
  \vdots        & \vdots \\
   \avrg{\beta_{u,n}^+} & \avrg{\beta_{d,n}^+} \\
  \avrg{\beta_{u,1}^-} & \avrg{\beta_{d,1}^-} \\
  \vdots        & \vdots \\
   \avrg{\beta_{u,n}^-} & \avrg{\beta_{d,n}^-} \\
 \end{array}
\right)
\left(
\begin{array}{cc}
\Delta u \\
\Delta d \\
\end{array}
\right) \, .
\end{eqnarray}
This leads to
\begin{eqnarray}
  \lefteqn{\mbox{cov}(\Delta u, \Delta d) = } \nonumber\\
 && \hspace{-1.3cm} \left(
 \begin{array}{cc}
 \sum_{i=1}^n N_i^+ \avrg{\beta_{u,i}^+}^2  + \sum_{i=1}^n N_u^- \avrg{\beta_{u,i}^-}^2  &  
        \sum_{i=1}^n N_i^+ \avrg{\beta_{u,i}^+} \avrg{\beta_{u,i}^+}  +\sum_{i=1}^n N_i^- \avrg{\beta_{u,i}^-} \avrg{\beta_{u,i}^-}\\
  \sum_{i=1}^n N_i^+ \avrg{\beta_{u,i}^+} \avrg{\beta_{u,i}^+}  +\sum_{i=1}^n N_i^- \avrg{\beta_{u,i}^-} \avrg{\beta_{u,i}^-} &  
  \sum_{i=1}^n N_i^+ \avrg{\beta_{d,i}^+}^2  + \sum_{i=1}^n N_d^- \avrg{\beta_{d,i}^-}^2 \\
 \end{array}
 \right) \, , \nonumber
 \end{eqnarray}
which equals the FOM of MLH and weighting method in the limit of infinite number of bins $n$.
For one bin one finds the FOM of the counting rate method.

\subsection{Comparison of FOM in different methods}

\begin{figure}
  \includegraphics[width=\textwidth]{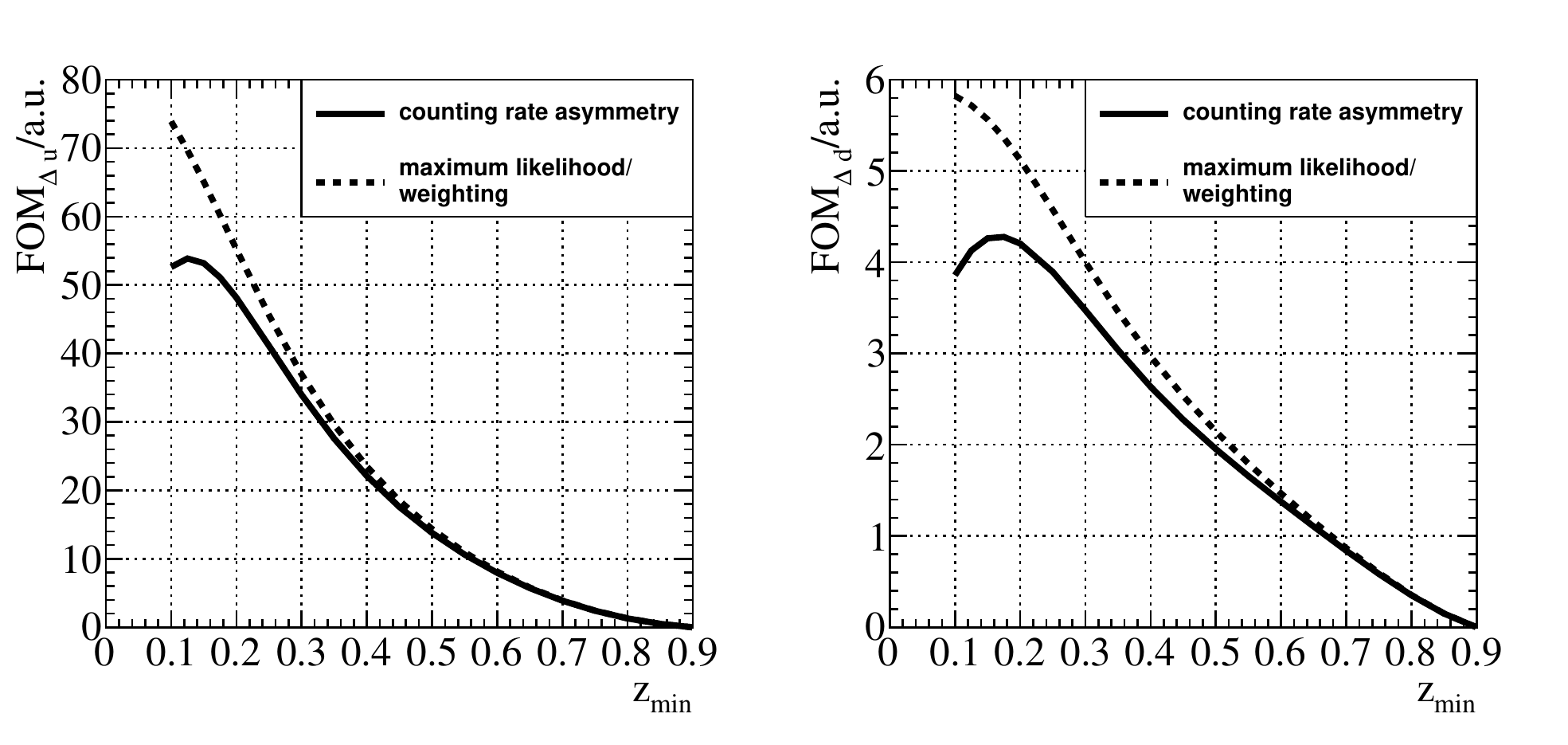}
  \caption{Comparison of FOM for $\Delta u$ (left) and $\Delta d$ (right) for
  the counting rate method (continuous line) and the Weighting/MLH method
  (dashed line) \label{fig:fom}}
 \end{figure}

Comparing eqs.~(\ref{eq:fom_w1}) and (\ref{eq:fom_mlh}) one realizes that 
the first one contains factors like $\avrg{\beta}^2$ as compared to
$\avrg{\beta^2}$ in the latter one.
Thus one expects the figure of merit to be larger for the maximum likelihood/weighting method as compared to the
counting rate method.
Fig.~\ref{fig:fom} shows a comparison of the FOM for $\Delta u$ and $\Delta d$ 
for the two methods using data in the interval $[z_{min},0.9]$.
  The figure of merit of the maximum likelihood/weighting method is always larger than the FOM of the counting rate method.
 Using only data with $z>0.2$ the gain is 15\% for $\Delta u$ and 22\% for $\Delta d$.
 Including data at lower $z$ the increase is 39\% and 37\% for $\Delta u$ and $\Delta d$, respectively.  
 In the counting rate method including data at low $z$ may even decrease the FOM  
 which clearly shows that this cannot be the optimal way to analyse the data.
 At large $z_{min}$ the FOM for both methods reach the same value 
 because $z$-dependence in a short $z$-interval is weaker as compared to a larger interval.
 Tab.~\ref{tab:fom} compares the different methods discussed in the paper.
 
 \begin{table}
\begin{center}
  \begin{tabular}{|l||l|l|l|l|}
 \hline
 Method & Counting Rates & MLH &  weighting &  binning in $z$ \\
 \hline \hline
 FOM     &   non-optimal        & optimal    &  optimal          & optimal for \\
         &                      &            &                   &$N_{bin} \rightarrow \infty$ \\
\hline
         drawbacks  &   \hspace{2mm} see above   $\ua$        &  CPU intensive,      &          &  empty bin   \\  
       &                &  correlation ($\pi^+, \pi^-)$&                  &    problem             \\
         &                &  hard to implement  &               &                 \\       
 \hline
 \end{tabular}
\end{center} 
 \caption{Comparison of the various methods discussed.\label{tab:fom}}
 \end{table}

 \section{Summary and Conclusions}
 In this paper a new method based on event weighting is proposed to extract quark helicity distributions with
 higher accuracy compared to the extraction based on counting rate asymmetries.
 Although the discussion was limited to two quark flavors and two asymmetries 
 the methods presented here can easily be extended to more quark flavors and asymmetries.
 The weighting method could also be applied to other parton distribution functions, e.g. transversity.
 In this paper only a leading order QCD analysis was discussed. In next-to-leading order the relation
 between helicity distributions and asymmetries is no more linear and it remains to be shown
 whether or how a weighting procedure can be implemented in this case.

 \section*{Acknowledgments}
 I would like to thank the Kavli Institute for Theoretical
 Physics and the Rice family fund
 who made an extended stay at the University of California Santa Barbara
 possible, during which this work was achieved. 
This research was  supported in part by the National Science Foundation under Grant No. PHY11-25915.
 I would like to thank Marcin Stolarski for discussions on the subject
and T.~W.~Donnelly for carefully reading the manusscript.

 \appendix
 \section{Covariance matrix for weighted asymmetries}\label{app1}
 In ref.~\cite{Pretz:2011qb} (appendix A) it is shown that the covariance between
 two weight factors $f$ and $g$ is given by
 \begin{equation}\label{cov_w1w2}
  \mbox{cov}(\sum f_i, \sum g_i) = N \avrg{fg} \approx \sum_i f_i g_i \, .
 \end{equation}
 Using this relation, the covariance matrix of the sums
 \[
{\boldsymbol \beta} = (\sum_{\ua\da} \beta_u^+, \sum_{\ua \ua} \beta_u^+,
  \sum_{\ua\da }(\beta_u^+)^2,\sum_{\ua \ua}(\beta_u^+)^2,
   \sum_{\ua \da } \beta_d^+, \sum_{\ua \ua} \beta_d^+,
\sum_{\ua \da } (\beta_d^+)^2, \sum_{\ua \ua} (\beta_d^+)^2  )
\] 
 entering the asymmetries $a_{\beta_u^+}$ and $a_{\beta_d^+}$
 can  be evaluated.
 
 To do this, it is convenient to define the following matrices 
 \[
  {\bf a_{q}} =
 \left(
  \begin{array}{cccc}
   N^+_{\ua \da} \avrg{(\beta_{q}^+)^2}_{\ua\da} & 0 & N^+_{\ua \da} \avrg{(\beta_q^+)^3}_{\ua\da} & 0 \\ 
   0 & N^+_{\ua \ua} \avrg{(\beta_q^+)^2}_{\ua\ua} & 0 & N^+_{\ua \ua} \avrg{(\beta_q^+)^3}_{\ua\ua} \\
 N^+_{\ua \da} \avrg{(\beta_q^+)^3}_{\ua\da} & 0 & N^+_{\ua \da} \avrg{(\beta_q^+)^4}_{\ua\da} & 0  \\ 
   0 & N^+_{\ua \ua} \avrg{(\beta_q^+)^3}_{\ua\ua} & 0 & N^+_{\ua \ua} \avrg{(\beta_q^+)^4}_{\ua\ua}      \\
\end{array}
\right)
\]
with $q=u,d$, and 
  \[
  {\bf b} =
 \left(
  \begin{array}{cccc}
   N^+_{\ua \da} \avrg{\beta_u^+ \beta_d^+}_{\ua\da}& 0 & 0 & 0  \\ 
   0 & N^+_{\ua \ua} \avrg{\beta_u^+ \beta_d^+}_{\ua\ua} &0&0\\
   0 & 0 & N^+_{\ua \da} \avrg{(\beta_u^+)^2 (\beta_d^+)^2}_{\ua\da}& 0 \\ 
   0 & 0 & 0 & N^+_{\ua \ua} \avrg{(\beta_u^+)^2 (\beta_d^+)^2}_{\ua\ua}     \\
\end{array}
\right) \, .
 \]
$\avrg{\dots}_{\ua\da (\ua\ua)}$ denote the expectation values
in the two spin configurations. 
 
The covariance matrix of $\boldsymbol \beta$ is given by
\[
 \mbox{cov}(\boldsymbol \beta) =
 \left(
 \begin{array}{cc}
  {\bf a_u} & {\bf b}  \\
  {\bf b} & {\bf a_d}
 \end{array}
\right) \, .
 \]
For the derivative of ${\bf a_{\beta}} = (\hat a_{\beta_u^+},\hat a_{\beta_d^+})$
with respect to ${\boldsymbol \beta}$ one finds
\[
\frac{\partial \bf {a_{\beta}}}{\partial {\boldsymbol \beta}} =
\left(
\begin{array}{cccccccc}
 \multicolumn{4}{c}{c_u} & 0 &0 &0 &0 \\
0& 0 & 0 & 0 & \multicolumn{4}{c}{c_d}  \\
\end{array}
\right)
\]
with
\[
c_q = \frac{1}{N^+ \avrg{(\beta_u^+)^2}} \, ( 1 , -1 , \hat
a_{\beta_q^+}, \hat a_{\beta_q^+} )
\]

Finally, neglecting terms proportional to the asymmetries $a_{\beta_q^+}$,
the upper left $2\times 2$ corner of the covariance matrix in eq.~(\ref{cov_aw}) is given by
\[
\frac{\partial \bf {a_{\beta}}}{\partial {\boldsymbol \beta}} \, \mbox{cov}({\boldsymbol \beta}) \, \frac{\partial \bf {a_{\beta}}}{\partial {\boldsymbol \beta}}^T \, .
\]
The lower left corner can be obtained analogously.
Note, if there is a correlation between the event rates of $\pi^+$ and
$\pi^-$ the lower left and upper right corner of
$\mbox{cov}(\bf a)$ in eq.~(\ref{cov_aw}) have non-zero entries as well.
These can be evaluated in the following way.
Correlations between positively and negatively charged pions 
arise when in one deep inelastic event a positively and a negatively charged 
hadron is observed.
The covariance for the corresponding sum of weights is, according to eq.~(\ref{cov_w1w2}), given by
\begin{eqnarray*}
\mbox{cov}(\sum \beta_u^+,\sum \beta_u^-) &=& \sum  \beta_u^+ \beta_u^-
 = N^{+ \mathrm{and} -} \, \avrg{ \beta_u^+ \beta_u^-} \, ,
\end{eqnarray*}
where the sum runs now over all deep inelastic events where at least one
positively {\em and} one negatively charged hadron is detected ($N^{+ \mathrm{and} -}  $).
The corresponding entry in the covariance matrix for the asymmetries in equation~(\ref{cov_aw}) is
\begin{eqnarray}
  \frac{\sum \beta_u^+ \beta_u^-}{\sum (\beta_u^+)^2 \sum (\beta_u^-)^2}
 &=&  \frac{N^{+ \mathrm{and} -}  \avrg{\beta_u^{+} \beta_u^-}_{+ \mathrm{and}
 -}}{N^+ N^- \, \avrg{(\beta_u^{+})^2 }_{+}\avrg{(\beta_u^{-})^2}}_{-} \, .\label{cov_plmi}
\end{eqnarray}
Care has to be taken in writing the left-hand side of equation~(\ref{cov_plmi}) in terms of expectation values.
The subscript "$+ \mathrm{and} -$" indicates that the expectation
value is taken with respect to all event where there was at least one
positively {\em and} one negatively charged hadron detected.
The subscripts "$+$ ($-$)" indicate that the  expectation
value is taken over all events with at least one 
positively (negatively) charged hadron.
Note that if several positively (or negatively) charged hadrons are detected
in an event, the corresponding event weight is the sum of all the weights.

  \bibliographystyle{IEEEtran}
 \bibliography{/home/pretz/bibtex/literature_dis.bib,/home/pretz/bibtex/statistics.bib,/home/pretz/bibtex/literature_edm}

\begin{thebibliography}{10}
\providecommand{\url}[1]{#1}
\csname url@samestyle\endcsname
\providecommand{\newblock}{\relax}
\providecommand{\bibinfo}[2]{#2}
\providecommand{\BIBentrySTDinterwordspacing}{\spaceskip=0pt\relax}
\providecommand{\BIBentryALTinterwordstretchfactor}{4}
\providecommand{\BIBentryALTinterwordspacing}{\spaceskip=\fontdimen2\font plus
\BIBentryALTinterwordstretchfactor\fontdimen3\font minus
  \fontdimen4\font\relax}
\providecommand{\BIBforeignlanguage}[2]{{%
\expandafter\ifx\csname l@#1\endcsname\relax
\typeout{** WARNING: IEEEtran.bst: No hyphenation pattern has been}%
\typeout{** loaded for the language `#1'. Using the pattern for}%
\typeout{** the default language instead.}%
\else
\language=\csname l@#1\endcsname
\fi
#2}}
\providecommand{\BIBdecl}{\relax}
\BIBdecl

\bibitem{Fernow:1981fw}
R.~C. Fernow and A.~D. Krisch, ``{High-energy Physics With Polarized Proton
  Beams},'' \emph{Ann. Rev. Nucl. Part. Sci.}, vol.~31, pp. 107--144, 1981.

\bibitem{Pretz:2011qb}
J.~Pretz, ``{Comparison of methods to extract an asymmetry parameter from
  data},'' \emph{Nucl. Instrum. Meth.}, vol. A659, pp. 456--461, 2011.

\bibitem{Pretz:2008mi}
J.~Pretz and J.-M. Le~Goff, ``{Simultaneous Determination of Signal and
  Background Asymmetries},'' \emph{Nucl. Instrum. Meth.}, vol. A602, pp.
  594--596, 2009.

\bibitem{Alekseev:2008cz}
M.~Alekseev \emph{et~al.}, ``{Direct Measurement of the Gluon Polarisation in
  the Nucleon via Charmed Meson Production},'' 2008.

\bibitem{Adolph:2015cvj}
C.~Adolph \emph{et~al.}, ``{Leading-order determination of the gluon
  polarisation using a novel method},'' 2015.

\bibitem{Alekseev:2010ub}
M.~G. Alekseev \emph{et~al.}, ``{Quark helicity distributions from longitudinal
  spin asymmetries in muon-proton and muon-deuteron scattering},'' \emph{Phys.
  Lett.}, vol. B693, pp. 227--235, 2010.

\bibitem{deFlorian:2007aj}
D.~de~Florian, R.~Sassot, and M.~Stratmann, ``{Global analysis of fragmentation
  functions for pions and kaons and their uncertainties},'' \emph{Phys. Rev.},
  vol. D75, p. 114010, 2007.

\bibitem{ffgen}
\BIBentryALTinterwordspacing
J.-P.~G. F.~Arleo. Fragmentation function generator. [Online]. Available:
  \url{http://lapth.cnrs.fr/ffgenerator}
\BIBentrySTDinterwordspacing

\bibitem{Barlow:213033}
\BIBentryALTinterwordspacing
R.~J. Barlow, \emph{{Statistics: a guide to the use of statistical methods in
  the physical sciences}}, ser. Manchester physics series.\hskip 1em plus 0.5em
  minus 0.4em\relax Chichester: Wiley, 1989. [Online]. Available:
  \url{https://cds.cern.ch/record/213033}
\BIBentrySTDinterwordspacing

\bibitem{Leader:2010rb}
E.~Leader, A.~V. Sidorov, and D.~B. Stamenov, ``{Determination of Polarized
  PDFs from a QCD Analysis of Inclusive and Semi-inclusive Deep Inelastic
  Scattering Data},'' \emph{Phys. Rev.}, vol. D82, p. 114018, 2010.

\bibitem{Arbabifar:2013tma}
F.~Arbabifar, A.~N. Khorramian, and M.~Soleymaninia, ``{QCD analysis of
  polarized DIS and the SIDIS asymmetry world data and light sea-quark
  decomposition},'' \emph{Phys. Rev.}, vol. D89, no.~3, p. 034006, 2014.

\end{thebibliography}

\end{document}